\documentclass[12pt]{article}
\usepackage{amsmath,amsfonts}
\makeatletter \@addtoreset{equation}{section}

\makeatother
\def\one{{\hbox{ 1\kern-.8mm l}}}
\newcommand{\Dslash}{\not{\hbox{\kern-4pt $D$}}}
\newcommand{\pdslash}{\not{\hbox{\kern-2pt $\partial$}}}
\newcommand{\be}{\begin{equation}}
\newcommand{\bea}{\begin{eqnarray}}
\newcommand{\eea}{\end{eqnarray}}
\newcommand{\ba}{\begin{array}}
\newcommand{\ea}{\end{array}}
\newcommand{\ee}{\end{equation}}
\newcommand{\nn}{\nonumber}

\begin{document}
\begin{titlepage}
 \hfill
 \vbox{
    \halign{#\hfil         \cr
           %hep-th/0411087 \cr
           %IPM/P-2004/066 \cr
           %SU-ITP-02/10 \cr
           } % end of \halign
      }  % end of \vbox
 \vspace*{20mm}
 \begin{center}
 {\Large {\bf On Geodesic Motion in Ho\v{r}ava-Lifshitz Gravity}\\ }

 \vspace*{15mm} \vspace*{1mm} { Amir Esmaeil Mosaffa\footnote{E-mail:mosaffa@physics.sharif.edu}}

 \vspace*{1cm}

 {\it Department of Physics, Sharif University of Technology \\
P.O. Box 11365-9161, Tehran, Iran}

 \vspace*{1cm}
 \end{center}

 \begin{abstract}
 We propose an action for a free particle in Ho\v{r}ava-Lifshitz gravity based on Foliation Preserving Diffeomorphisms. The action reduces to the usual relativistic action in the low energy limit and allows for subluminal and superluminal motions with upper and lower bounds on velocity respectively. We find that deviation from general relativity is governed by a position dependent coupling constant which also depends on the mass of the particle. As a result,  light-like geodesics are not affected whereas massive particles follow geodesics that become mass dependent and hence  the equivalence principle is violated. We make an exact study for geodesics in flat space and a qualitative analysis for those in a spherically symmetric curved background.

 \end{abstract}

 \vspace{2cm}
 \begin{center}
 \it{Dedicated to Farhad Ardalan on his 70th birthday}
 \end{center}
 \end{titlepage}

%%%%%%%%%%%%%%%%%%%%%%%%%%%%%%%%%%%%%%%%%%%%%%%%%%%%%%

\section{Introduction and Setup}
The long quest for unification remains an outstanding challenge for theoretical physics despite impressive progress in this regard in the last few decades. The one last step in this program is to unify quantum theory with gravity. Gravitational interactions, as predicted by Einstein's theory, turn out to be irrelevant such that a naive quantization of general relativity breaks down at high energies. A possible resolution for this problem is to add covariant higher derivative terms to Einstein-Hilbert action such that the superficial degree of divergence decreases. A new problem arises though because the higher temporal derivatives introduce ghost fields in the theory. 

A new idea in this regard has been to add only higher spatial derivatives to the Einstein-Hilbert action. By doing this, and at the expense of breaking general covariance, no ghost fields appear and at the same time the higher derivatives render the theory renormalizable. Inspired by similar ideas in Lifshitz's scalar field theory, and based on some earlier works \cite{Horava:2008jf,Horava:2008ih}, Ho\v{r}ava introduced a renormalizable theory for gravity  in \cite{Horava:2009uw} which has been named Ho\v{r}ava-Lifshitz (HL) gravity. This theory has attracted a lot of interest since then and its various aspects have been studied, such as finding solutions, addressing fundamental issues and introducing modifications in the theory and studying its cosmoligical consequences. Some of these works are listed in \cite{Lu:2009em,Horava:2009if,Kiritsis:2009sh,Blas:2009yd}.

A basic question in this theory is to understand how matter fields influence geometry, that is, in this new theory for gravitation, what is the analogue of the matter energy-momentum tensor that acts as a gravitational source in Einstein's field equations in general relativity. A first step to answer this question is to study how particles move in a given background or in other words what are the geodesics in the HL theory.

The above question has been addressed in some works recently\cite{Chen:2009bu,Capasso:2009fh,Rama:2009xc}. Among these \cite{Capasso:2009fh} makes a comprehensive study of particles in HL gravity as the optical limit of a scalar field theory. They find new features such as superluminal motions and massive luminal particles as well as the dependence of geodesics on the mass of the particle. Similar results are found in \cite{Rama:2009xc} by studying super Hamiltonian formalism.

In this note we study geodesic motion for free particles in the same spirit that HL theory is built itself. We write down a kinetic action compatible with the symmetries of the theory and deform it in a way that it reproduces the expected relativistic action in the low energy limit. 

Our results share some of the features with the mentioned works; it predicts sub and superluminal motions as two different sectors for motion. In the former sector, we find an upper bound on the velocity, quite similar to relativistic dynamics, whereas in the latter we find a lower bound on the velocity. We also find that even in the subluminal sector a particle's path in general deviates from its relativistic counterpart. The deviation is in terms of a coupling constant which depends on the mass of the particle and is in general position dependent. Massless particles, on the other hand, move exactly as predicted by general relativity.

In the remainder of this introduction  we briefly list the basic ingredients of the  HL gravity which we need in the rest of the work. 
\newpage
We consider a $D+1$ space-time
with the metric written in the $ADM$ form as
%We want to study the motion of free (massive) particles in the gravitational background provided by the
%Horava-Lifshitz theory. We start with a microscopic action for the particle compatible with the symmetries of
%the theory and then add relevant deformations such that we end up with an accidentally enhanced
%symmetry in the infrared. As symmetries we start with are not as large as diffeomorphisms, one might have several options for the microscopic action as well as deformations.

%\subsection*{Setup}

\be
ds^2=-N^2 dt^2+g_{ij}\  du^idu^j\nonumber
\ee
where
\be
du^i=dx^i+N^idt\, ,\ \ \ \ \ \ \ \ i=1,2,...,D\nonumber
\ee
and $N$ and $N^i$ are the usual lapse and shift functions and $g_{ij}$ is the metric on the spatial section.
HL gravity is based on the general coordinate transformations which preserve the foliation of space-time into the time direction and spatial sections and thus the name
``foliation preserving diffeomorphisms" or FPD in short (one way of identifying these is to make a nonrelativistic contraction of diffeomorphisms). FPD in its finite form turns out to be
\be
t\rightarrow t'(t) ,\ \ \ \ \ x^i\rightarrow x'^i(t,\vec{x})\nonumber
\ee
under which
\bea
g_{ij}&\rightarrow& g'_{ij}=g_{kl}\ \frac{\partial x^k}{\partial x'^i}\ \frac{\partial x^l}{\partial x'^j}\nn\\
N_i&\rightarrow& N'_i=g_{kl}\ \frac{\partial x^k}{\partial x'^i} \ \frac{\partial x^l}{\partial t'}+
N_k\  \frac{\partial x^k}{\partial x'^i}\ \frac{d t}{d t'}\nn\\
N&\rightarrow& N'=N\ \frac{d t}{d t'}\nn
\eea
FPD allows for more invariants than diffeomorphisms. For example for any two spacetime vectors $A^\mu$ and $B^\nu$
one can form the following invariant products
\be
N^2A^tB^t\ ,\ \ \ g_{ij} (A^i+N^iA^t)(B^j+N^jB^t)\ ,\ \ \ \frac{1}{N^2}(A_t-N^iA_i)(B_t-N^jB_t)\ ,\ \ \
g^{ij}A_iB_j\nn
\ee
Apart from FPD, the HL gravity assumes that the microscopic theory enjoys a scaling symmetry that acts anisotropically on space and time. This symmetry is characterized by an exponent $z>1$ and acts as follows
\be
x^i\rightarrow bx^i\ ,\ \ \ \ \ t\rightarrow b^z t\nn
\ee
This means that associating a scaling dimension $-1$  to $x^i$, time will have scaling dimension $-z$.
We now move on to study the motion of particles in this theory.
%%%%%%%%%%%%%%%%%%%%%%%%%%%%%%%%%%%%%%%%%%%%%%%%%%%%%%%%%%%%%%%%%%%%%%%%

\section{Particles in HL gravity}

We want to write down an action for a particle that exhibits FPD symmetry at all regimes of energy. In the UV the action must have an anisotropic scaling symmetry between time and space, specified by a critical exponent $z$, and in the IR it should roll down to a relativistic theory. The microscopic action which governs the UV regime consists of kinetic and potential terms. As for the kinetic term, the critical exponent only shows up in the scaling dimension of the coupling constant, that is, the form of the kinetic term is not sensitive to $z$. The form of the potential terms, however, is severely restricted by the value of $z$.  In the following we study particles that move freely and hence feel no potential in the UV.  Such a UV action is most naturally written as

\be\label{kinetic}
S_{UV}\sim\int d\tau\ g_{ij}\ \dot{u}^i\dot{u}^j
\ee
where $\tau$ is an invariant of FPD and "dot" means derivative with respect to $\tau$. It is natural to choose time to parameterize the world line of the particle and hence set $d\tau=Ndt$. By this choice the dynamical fields will be $x^i(t)$ which define a valid set of degrees of freedom required to specify a curve in spacetime.  This choice of the parameter can be incorporated in the action by writing the following
\be\label{SUV}
S_{UV}\sim \frac{1}{2}\int d\tau \left[\frac{1}{e}\ g_{ij}\ \dot{u}^i\dot{u}^j+
\frac{e}{N^2\dot{t}^2}\ g_{ij}\ \dot{u}^i\dot{u}^j\right]
\ee
where $e(\tau)$ is the worldline einbein. Note that we now have $x^i(\tau)$, $t(\tau)$ and $e(\tau)$ as the degrees of freedom, two more than required. In return we have the equation of motion for $e$ and a new symmetry, reparameterization invariance, which is defined by 

\be\label{invar}
\tau\rightarrow \tau'(\tau)\ ,\ \ \ e(\tau)\rightarrow e'(\tau')=e(\tau) \frac{d\tau}{d\tau'}\ ,
\ \ \ (x^i,t)\rightarrow (x^i,t)
\ee
One can now use the reparametrization invariance to fix $e$ at a constant value, say $1$. The equation of motion for $e$ 
\be\label{para}
e^2= N^2\dot{t}^2
\ee
then fixes the parameter as $d\tau=Ndt$ and the resulting action will read
\be\label{suv1}
S_{UV}\sim\int \frac{dt}{N}\  g_{ij}\ (\frac{dx^i}{dt}+N^i)(\frac{dx^j}{dt}+N^j)
\ee

 As for the IR action, we want it to be of the form

 \be\label{SIR} 
 S_{IR}\sim\int d\tau\sqrt{c^2N^2\dot{t}^2-g_{ij}\dot{u}^i\dot{u}^j}
 \ee
 where now $\tau$ is a full diffeomorphism invariant parameter and $c$ is the speed of light. One can write a quadratic form of this action by introducing the world line einbein 
  \be\label{quad}
S_{IR}\sim\int d\tau \left[\frac{1}{e}(g_{ij}\ \dot{u}^i\dot{u}^j-c^2N^2\dot{t}^2)- e\right]
\ee
where again we have the same symmetry of (\ref{invar}). The same procedure stated above is applicable to (\ref{quad}); use (\ref{invar}) to fix $e$, use equation of motion for $e$ to
fix $d\tau^2\sim  c^2N^2\dot{t}^2-g_{ij}\dot{u}^i\dot{u}^j$ and the resulting action will be the familiar relativistic action for a free particle.

Now let's put everything together, we start with the microscopic action (\ref{SUV}) and, in order to end up with (\ref{quad}) in the IR, we add to it  a deformation of the form
\be
S_{def}\sim\int d\tau \left[\frac{1}{e}c^2N^2\dot{t}^2+e\right]
\ee
so that the final form for the action will be $S_{UV}+S_{def}$. One should insert various coupling constants to have the correct dimensions for the fields. Given all this,  we propose the following action for the motion of a free particle in the HL gravity 
\be\label{action}
S=\frac{1}{2}\int d\tau \left[\frac{1}{e}\ (g_{ij}\ \dot{u}^i\dot{u}^j-c^2N^2\dot{t}^2)+
\frac{eM^2}{N^2\dot{t}^2}\ (g_{ij}\ \dot{u}^i\dot{u}^j-\frac{m^2}{M^2}c^2N^2\dot{t}^2)\right]
\ee
In the following we will relate the coupling constants $m$ and $M$ to  physical quantities. The engineering dimensions of the fields and constants are as follows
\be
[x]=-1\ ,\ \ \ [t]=[\tau]=-z\ ,\ \ \ [m]=[M]=[e^{-1}]=2-z\ ,\ \ \ [c]=z-1
\ee

To identify the physical characteristics of the particle in terms of the coupling constants we consider the case of flat space, that is, when $N=1$ and $N^i=0$. The equation of motion for $e$ gives
\be\label{eomein}
e^2M^2= \dot{t}^2\ \frac{\dot{x}^i\dot{x}_i-c^2\dot{t}^2}{\dot{x}^i\dot{x}_i-\frac{m^2}{M^2} c^2\dot{t}^2}\ ,\ \ \ \ \ \dot{x}^i\dot{x}_i=g_{ij}\ \dot{x}^i\dot{x}^j
\ee
One immediately identifies two distinct cases; the numerator and denominator in the above expression are either both positive or both negative. In the former case one faces superluminal and in the latter subluminal motions. Let us first consider the subluminal motions. Substituting $e$ from (\ref{eomein}) in (\ref{action}) and after some algebra we arrive at
%\be\label{sub1}
%S_{SUB}=-\frac{\alpha}{\lambda}\int d\tau\ \sqrt{c^2\dot{t}^2-\dot{x}^i\dot{x}_i}\ \sqrt{1-\frac{\lambda^2}{\alpha^2k^4}\frac{dx^i}{dt}\frac{dx_i}{dt}}
%\ee
%One can choose $\tau=t$ in the above expression to give
\be\label{sub2}
S_{SUB}=-mc^2\int dt\ \sqrt{1-\frac{\dot{x}^i\dot{x}_i}{c^2}}\ \sqrt{1-{\left(\frac{M}{m}\right)}^2\frac{\dot{x}^i\dot{x}_i}{c^2}}\ ,\ \ \ \ \ \  \dot{x}^i=\frac{dx^i}{dt} 
\ee
If we want a relativistic point particle action in the IR, we need the second square root in the above expression to capture corrections to the relativistic action. This requires the second square root to be almost equal to one even when the particle's velocity is comparable to the speed of light. A minimum requirement for this is to assume $\frac{m}{M}> 1$. The larger $\frac{m}{M}$ is, the better we can approximate the IR motion with a relativistic action. 
%Given that special relativity is valid to excellent precision even at high sub luminal velocities, it is plausible to assume that
%\be
%\frac{\alpha}{\lambda}k^2\gg c
%\ee
A second consequence of (\ref{sub2}) is to identify $m$ with the rest mass of the particle.

Now let's move to superluminal motions. This is when both the numerator and the denominator in (\ref{eomein}) are positive. Again substituting $e$ from (\ref{eomein}) in (\ref{action}) gives
\be\label{super}
S_{SUP}=M\int dt\ \dot{x}^i\dot{x}_i\ \sqrt{1-\frac{c^2}{\dot{x}^i\dot{x}_i}}\ \sqrt{1-{\left(\frac{m}{M}\right)}^2\frac{c^2}{\dot{x}^i\dot{x}_i}}\ ,\ \ \ \ \ \  \dot{x}^i=\frac{dx^i}{dt} 
\ee
For large velocities the above action approximates (\ref{suv1}) in flat space.  $M$ has the same dimension as $m$ so we identify it with some sort of a ``mass" in the superluminal regime.

Collecting everything together, the action (\ref{action}), in the case of a flat space, allows sub and superluminal motions for a particle. The subluminal motion is governed by relativistic action plus corrections
\be
S_{SUB}=-mc^2\int dt \ \sqrt{1-\frac{\dot{x}^i\dot{x}_i}{c^2}}\ \left[1-\frac{1}{2}\  g^2 \ \frac{\dot{x}^i\dot{x}_i}{c^2}+\cdots\right]
\ee 
whereas the superluminal motion is governed by the non relativistic  action for the particle plus corrections
\be
S_{SUP}=M\int dt\ \dot{x}^i\dot{x}_i \left[ 1-\frac{1}{2} \frac{c^2}{\dot{x}^i\dot{x}_i}+\cdots\right]\ \left[1-\frac{1}{2}\frac{1}{g^2} \frac{c^2}{\dot{x}^i\dot{x}_i}+\cdots\right]
\ee
where 
\be
g=\frac{M}{m}
\ee

Some comments are in order
\begin{itemize}
\item Of the three constants appearing in the gauge fixed action, $c$ is the emergent speed of light and is provided by the gravitational part of the HL theory whereas $m$ and $M$ 
are physical properties of the particle under study.
\item For a given set of constants, $c$, $m$ and $M$, not all velocities are allowed for a particle. Either there is an upper bound on the velocities, $c$, or a lower bound  $c/g$. The initial condition of the motion determines whether the motion is subluminal or superluminal.
\item Given the excellent validity of special relativity even at velocities close to the speed of light, the action $S_{SUB}$ must give a small correction to the relativistic action. We should thus assume that $g\ll 1$ or $m\gg M$ for usual particles.
\item For $g\ll 1$, the action $S_{SUP}$, having an expansion in $1/g$ becomes ``strongly coupled" even at very high velocities and is approximated by $S_{UV}$ in a very narrow range of velocities.
%\footnote{It is unclear to the author whether this has anything to do with a similar observation in \cite{} regarding the ``extra mode" of the HL gravity.}. 
\end{itemize}

Now that we have identified the coupling constants and also realized the necessity of the assumption $g\ll1$, we move on to the general case of a curved background. Here again the action (\ref{action}) predicts sub and superluminal motions. The former is governed by the usual relativistic action from general gravity plus corrections
\bea\label{SUBper}
S_{SUB}&=&-mc^2\int dtN\ \sqrt{1-\frac{1}{N^2}\frac{\dot{u}^i\dot{u}_i}{c^2}}\ \sqrt{1-\frac{g^2}{N^2}\frac{\dot{u}^i\dot{u}_i}{c^2}}\cr &&\cr
&=&\int dt \ \mathcal{L}_{GR}\  \left[1-\frac{1}{2}\frac{g^2}{c^2N^2}(\dot{x}^i+N^i)(\dot{x}_i+N_i)+\cdots\right]
\eea
whereas the superluminal motion is governed by the non relativistic action plus corrections
\bea\label{SUPper}
S_{SUP}&=&M\int dt\frac{1}{N} \dot{u}^i\dot{u}_i\sqrt{1-N^2\frac{c^2}{\dot{u}^i\dot{u}_i}}\sqrt{1-\frac{N^2}{g^2}\frac{c^2}{\dot{u}^i\dot{u}_i}}\cr &&\cr
&=&=\int dt\ \mathcal{L}_{NR}\ \left[1-\frac{1}{2}N^2\frac{c^2}{\dot{u}^i\dot{u}_i}\cdots\right]\ \left[1-\frac{1}{2}\frac{N^2}{g^2}\frac{c^2}{\dot{u}^i\dot{u}_i}+\cdots\right]
\eea
where
\be
\mathcal{L}_{GR}=-mc^2N\ \sqrt{1-\frac{1}{N^2}\frac{\dot{u}^i\dot{u}_i}{c^2}}\ ,\ \ \ \ \ \mathcal{L}_{NR}=\frac{M}{N} \dot{u}^i\dot{u}_i\ ,\ \ \ \ \  \dot{u}^i=\frac{dx^i}{dt}+N^i 
\ee
In the next section we study the equations of motion following from (\ref{action}) to find geodesics in some simple backgrounds.

%%%%%%%%%%%%%%%%%%%%%%%%%%%%%%%%%%%%%%%%%%%%%%%%%%%%%%%%%%%%%%%%%%%%%%%%%%
%%%%%%%%%%%%%%%%%%%%%%%%%%%%%%%%%%%%%%%%%%%%%%%%%%%%%%%%%%%%%%%%%%%%%%%%%%
\section{Geodesics in HL Gravity}
In what follows we assume that the constants $m$ and $M$ are finite and that $g=M/m\ll1$. To find the geodesics of a particle from the action (\ref{action}), we first write down the equation of motion for $e$
\be\label{Eoem}
e^2m^2=N^2\dot{t}^2\ \frac{\dot{u}^i\dot{u}_i-c^2N^2\dot{t}^2}{g^2\dot{u}^i\dot{u}_i-c^2N^2\dot{t}^2}
%\frac{1}{E}(\dot{u}^i\dot{u}_i-c^2N^2\dot{t}^2)=\frac{EM^2}{N^2\dot{t}^2}(\dot{u}^i\dot{u}_i-\frac {c^2}{g^2}N^2\dot{t}^2)
\ee
This equation specifies the allowed range of velocities and the qualitative characteristics of the motion. There are as usual two possible routes to find the geodesics. One is to solve $e$ from above, replace in the original action (\ref{action}), and solve the equations of motion following from the resulting action (\ref{SUBper})/(\ref{SUPper}). The second route is to use reparametrization invariance to fix $e$ at a suitable constant which one can choose to be $e=1/m^2$. Replacing this in (\ref{action}) gives
\be\label{gfaction}
S_{GF}=\frac{m}{2}\int d\tau\left[\dot{u}^i\dot{u}_i-c^2N^2\dot{t}^2+g^2\frac{\dot{u}^i\dot{u}_i}{N^2\dot{t}^2}\right]
\ee
There is also a constraint which comes from gauge fixing
\be\label{cons}
 \dot{u}^i\dot{u}_i-c^2N^2\dot{t}^2=g^2\frac{\dot{u}^i\dot{u}_i}{N^2\dot{t}^2}-c^2
 \ee
 To find the geodesics we should solve the equations of motion coming from (\ref{gfaction}) together with the constraint (\ref{cons}) that specifies the world line parameter. Note that for $g=0$, the above system of equations is exactly that in general relativity for geodesic motion.
 In the following we first solve this system of equations for flat space. In the passing, we study some limiting velocities and masses including the zero mass limit. We then make an approximate analysis for a spherically symmetric background to examine some qualitative features of particle motion. 
 
 %\be
%S=-mc\int d\tau\ \sqrt{c^2N^2\dot{t}^2-\dot{u}^i\dot{u}_i}\ \sqrt{1-g^2\frac{\dot{u}^i\dot{u}_i}{c^2N^2\dot{t}^2}}
%\ee
%Note that the parameter $\tau$ is now at our disposal and can be chosen arbitrarily. In particular one may choose $d\tau=dt$
%\be
%S=-mc^2\int dt N\ \sqrt{1-\frac{\dot{u}^i\dot{u}_i}{c^2}}\ \sqrt{1-g^2\frac{\dot{u}^i\dot{u}_i}{c^2}}\ ,\ \ \ \ \ \dot{u}^i=\frac{dx^i}{dt}+N^i
%\ee
%Now one can expand in powers of $g^2$ to find corrections to relativistic equations perturbatively
%\be
%S=-mc^2\int dt N\  \sqrt{1-\frac{\dot{u}^i\dot{u}_i}{c^2}}\  [1- \frac{1}{2} g^2\frac{\dot{u}^i\dot{u}_i}{c^2}\cdots]
%\ee
%It is obvious that the leading term gives the relativistic action for the particle. The corrections include terms which depend on the velocities as well as the mass of the particle. The important conclusion is that the geodesics will depend on masses and equivalence principle will no longer hold\footnote {agreement with poly}. This is of course expected because, by construction, diffeomorphism is not a symmetry of our action.

%In the following we first apply the above results to flat space which is easy enough to allow for a nonperturbative analysis in both sub and super luminal cases. In the passing, we study some limiting velocities and masses including the zero mass limit. We then make an approximate analysis for a spherically symmetric background to obtain examine some qualitative features of particle motion.

\subsection{Flat Space}
Set $N=1$ and $N^i=0$. $S_{GF}$ results in constant $\dot{x}^i$ and $\dot{t}$. This gives a straight line in space with a constant velocity. Using spatial rotations we project the motion on the axis $x$. The constraint equation determines the range of allowed velocities and possible lower and upper bounds. So we have
%\be
%\dot{t}=a=constant,\ \ \ \ \dot{x}=b=constant\ ,\ \ \ \ \frac{\dot{x}}{\dot{t}}=\frac{b}{a}=v=constant
%\ee
%and the constraint
\be
1=a^2\frac{v^2/c^2-1}{v^2/c^2-1/g^2}\ ,\ \ \ \ \ v=\frac{dx}{dt}
\ee
The following situations can be identified
\begin{itemize}
\item $0< a<1$ and $1/g^2< v^2/c^2$. This is superluminal motion. In the limit  $a\rightarrow 1^{-}$, the velocity goes to infinity.
\item $1<a<1/g$. This is not an allowed initial condition.
\item $a\ge 1/g$ and $0\le v^2/c^2< 1$. This is subluminal motion.
\end{itemize}

{\bf{Limiting velocities and massless case}}\\
\\
There are some special limits on the masses which allow the particle to assume limiting velocities, the lower and upper bounds in the super and subluminal sectors respectively. To study these cases  we go back to the equation of motion for $e$ before fixing the gauge
\be\label{masses}
e^2=\frac{a^2}{m^2}\ \frac{v^2/c^2-1}{g^2v^2/c^2-1}
\ee
As the first limit we consider $g\rightarrow0$ with $m$ kept fixed. In this limit  the above relation becomes the usual relativistic constraint equation. All that remains is the subluminal motion and the theory is just special relativity. Obviously one can then take the $m=0$ limit which gives $v=c$. This is of course the upper bound for velocity in the subluminal sector.

The other special case is when $m\rightarrow\infty$ and $g$ is kept fixed and still smaller than one. In this limit the denominator in 
(\ref{masses}) should become zero, i.e., $v/c=1/g$, which is the lower bound for velocity in the superluminal sector.

\subsection{Spherically Symmetric Background}
We study geodesic motion in the Schwarzschild solution. This is not an exact solution of HL theory but is an approximation at sufficiently large distances from the origin.  Here we are only interested in general features of the geodesic motion and postpone a detailed analysis to further works. The following results should specially be considered with care when talking about regions near the horizon.

Start with the following metric 
\be
ds^2=-(1-R/r)dt^2+\frac{dr^2}{1-R/r}+r^2d\Omega^2_{d-1}
\ee
where $R$ is a positive constant and $\Omega$ is the transverse sphere. Because of spherical symmetry we can have purely radial motions and that is what we consider below. 
Far from the origin where $N^2=1-R/r\sim1$ we effectively have a flat space and the above arguments about flat space apply. Choosing the initial value then decides whether we start in the sub or superluminal sectors. Ay any given value of radius, the upper and lower bounds of velocity are determined by the position dependent coupling constant, $g/N^2$.  In terms of the radial velocity, $v=dr/dt$, the constraint equation is written as
\be\label{gheyd}
1=\frac{N^2 \dot{t}^2}{g^2}\ \frac{v^2/c^2-N^4}{v^2/c^2-N^4/g^2}
\ee
The first thing to note is that as long as $N^2$ does not change sign during the course of motion (the particle does not cross any horizons), once we start in a sector we remain in that all through the motion. In the subluminal sector the upper bound is now shifted from $c$ to $cN^2$ whereas in the superluminal sector the lower bound has shifted from $c/g$ to $cN^2/g$. As the particle moves towards the origin, both the upper and lower bounds decrease in size. For subluminal motion this results in the familiar fact that it takes infinite coordinate time for the particle to reach the horizon. In the superluminal sector this means that the particle need not necessarily slow down towards the origin. 

One may try to solve equations for the subluminal sector to compare with the general relativistic geodesics. For $g\ll1$ this can be done perturbatively. One can either use the action (\ref{SUBper}) directly or solve  (\ref{gheyd})  together with
\be
c^2N^2\dot{t}+g^2\frac{v^2}{N^4\dot{t}^2}=a
\ee
where $a$ is a constant and $v<cN^2$. Either method shows that the corrections to general relativistic equations of motion depend on the mass of the particle through $g$, that is, particles with different masses follow different paths in space. This violates the equivalence principle as expected. 

Interestingly for a massless particle for which $g$  has to be zero, no superluminal sector exists and the geodesics are those predicted by general relativity. If one defines light by such an assumption then its motion is the same in both Einstein and HL theories for gravity.

\section{Conclusions}
In this note we have proposed an action to describe particles' motions in HL gravity. The guiding principle in our construction is the fundamental symmetry of the theory, FPD, and the requirement that general relativity is reproduced in low energies. Starting with a non-relativistic action in UV, we add deformations such that the dominant terms in low energies sum up to the familiar relativistic action for particle motion. In doing so we encounter various coupling constants that are fixed by considering the flat space case.

The action allows for two distinct sectors for motion, one where velocities have to be smaller than an upper bound and the other which imposes a lower bound on velocities. We call the first sector subluminal and the second superluminal. 

Focusing on subluminal motions in flat space, we can describe the motion in terms of the particle's mass, $m$, and a coupling constant $g$. Insisting on the validity of special relativity even at velocities close to that of light, we have to assume that $g\ll1$. This allows us to write the correction to special relativity, imposed by HL theory, in powers of $g$.

For superluminal motions in flat space, the theory is most naturally written in terms of $g$ and a constant with mass dimension $M$. The assumption $g\ll1$ results in the fact that the relevant action for this sector becomes strongly coupled for arbitrarily large velocities. Therefore the non-relativistic action that we started with  is valid in an arbitrarily small range of velocities.

We also considered some limiting cases of the constants and velocities. We find that for $g=0$ the superluminal sector disappears. A particle with $m=0$ is then shown to travel at the upper bound velocity.  The other limiting velocity, the lower bound in superluminal sector, is reached when $M$ goes to infinity.

One might adopt a different point of view, i.e., require that the deformations to the microscopic action  introduce perturbative corrections to the non-relativistic action. This requries to assume that $g\gg1$ and results in consequences opposite to the above; the subluminal sector becomes strongly coupled at arbitrarily low energies and effectively special relativity never becomes a good approximation\footnote{It is a curiosity that this behavior is quite similar to an observation made in \cite{Blas:2009yd} regarding the ``extra mode" in HL theory.}.

For curved backgrounds the above features remain valid. The coupling constant now depends on position and since it also depends on the mass of the particle, the corrections to the relativistic action become mass dependent. As a result particles with different masses follow different geodesics and the equivalence principle is violated. Deviations from general relativistic geodesics are measured in terms of $g$. An interesting result is that light, for which $m=g=0$, is insensitive to the HL corrections to general relativity and still follows the usual light-like geodesics.

Solving the equations of motion of our action to find the actual geodesics can in general be difficult. In the subluminal sector, however, one can do this perturbatively. In this note we studied the flat space geodesics in both sectors and pointed out some qualitative features of geodesics in a spherically symmetric background in its subluminal sector. Exact calculations in this regard, and more general backgrounds, are postponed to future works.

Formulating the geodesic motion in the Lagrangian language, our approach may be taken as a first step towards an understanding of how matter behaves as a source in HL theory. This can be achieved by considering a system of the HL gravity plus the particle actions and treating the latter as a source to gravity. This is also postponed to future works.

\section*{Acknowledgments}
It is a pleasure to thank F.Ardalan and B.Lynn for useful discussions and M.Alishahiha for useful discussions and also comments on the draft. My sincere thanks are due to L.Alvarez-Gaume for posing the problem, very illuminating discussions and for a reading of the draft.   I would also like to thank the theory division of CERN for hospitality where part of this work was done.

%---------------------------------------------------------------------
%Bibliography

%--------------------------------------------------------------------
\end{document}